\documentclass[runningheads]{svmult}

\usepackage{makeidx}   
\usepackage{graphicx}  
\usepackage{subeqnar}  
\usepackage{multicol}  
\usepackage{physprbb}  
\makeindex             

\begin{document}
\title*{Jeans solutions for triaxial galaxies}
%
%
%
%
%
\author{Glenn~van~de~Ven\inst{1}
\and Chris~Hunter\inst{2}
\and Ellen~Verolme\inst{1}
\and Tim~de~Zeeuw\inst{1}}
\authorrunning{Van de Ven et al.}
%
%
\institute{Sterrewacht Leiden, Postbus 9513, 2300 RA Leiden, The
  Netherlands 
\and Department of Mathematics, Florida State University, Tallahassee,
FL 32306-4510} 

\maketitle              

\begin{abstract}
The Jeans equations relate the second-order velocity moments 
to the density and potential of a stellar system.     
For general three-dimensional stellar systems, there are three
equations, but these are not very helpful, as they contain six
independent moments.  
By assuming that the potential is triaxial and of separable St\"ackel
form, the mixed moments vanish in confocal ellipsoidal
coordinates. 
The three Jeans equations and three remaining non-vanishing moments
form a closed system of three highly-symmetric coupled first-order
partial differential equations in three variables. 
They were first derived by Lynden--Bell in 1960, but have resisted
solution by standard methods.   
Here we present the general solution by superposition of singular
solutions.
\end{abstract}


\section{Introduction}
\label{sec:introduction}

Much has been learned about the mass distribution and internal
dynamics of galaxies by modeling their observed kinematics with
solutions of the Jeans equations 
(e.g. \cite{1987gady.book.....B}). 
The Jeans equations connect the second-order velocity moments 
(or the velocity dispersions, if the mean streaming motion is known) 
directly to the density and the gravitational potential of the galaxy,
without the need to know the phase-space distribution function $f$. 
In nearly all cases there are fewer Jeans equations than velocity
moments, so that additional assumptions have to be made about the
degree of anisotropy. 
Furthermore, the resulting second moments may not correspond to a
physical distribution function $f\geq 0$. 
These significant drawbacks have not prevented wide application of the
Jeans approach to the kinematics of spherical and axisymmetric
galaxies. Many (components of) galaxies have triaxial shapes  
(\cite{1976MNRAS.177...19B}, 
\cite{1978MNRAS.183..501B}), 
including early-type bulges, bars, and giant elliptical galaxies. 
In this geometry, there are three Jeans equations, but little use has
been made of them, as they contain six independent second moments,
three of which have to be chosen ad-hoc 
(see e.g. \cite{2000MNRAS.318.1131E}).

An exception is provided by the special set of triaxial mass models
that have a gravitational potential of St\"ackel form. 
In these systems, the Hamilton--Jacobi equation separates in
confocal ellipsoidal coordinates 
(\cite{histStackel1891}), 
so that all orbits have three exact integrals of motion, which are
quadratic in the velocities.
The three mixed second-order velocity moments vanish, so that the
three Jeans equations for the three remaining second moments form a
closed system.   
Lynden--Bell (\cite{thesisLynden-Bell}) 
was the first to derive the explicit form of these Jeans equations.
He showed that they constitute a highly symmetric set of three
first-order partial differential equations for three unknowns,
each of which is a function of the ellipsoidal coordinates,
but he did not derive solutions. 

When it was realized that the orbital structure in the triaxial
St\"ackel models is very similar to that in numerical models
for triaxial galaxies with cores  
(\cite{1985MNRAS.216..273D}, 
 \cite{1979ApJ...232..236S}), 
interest in the second moments increased, and the Jeans equations were
solved for a number of special cases.    
These include the axisymmetric limits and elliptic discs 
(\cite{1988ApJ...333...90D}, 
\cite{1989MNRAS.236..801E}), 
triaxial galaxies with only thin tube orbits 
(\cite{1992ApJ...389...79H}), 
and the scale-free limit 
(\cite{2000MNRAS.318.1131E}).  
In all these cases the equations simplify to a two-dimensional
problem, which can be solved with standard techniques after
transforming two first-order equations into a single second-order
equation in one dependent variable.  
However, these techniques do not carry over to a single third-order
equation in one dependent variable, which is the best that one could
expect to have in the general case. 
As a result, the latter has remained unsolved.

We have solved the two-dimensional case with an alternative solution
method, which does not use the standard approach, but instead 
uses superposition of singular solutions.  
This approach can be extended to three dimensions, and provides the
general solution for the triaxial case in closed form. 
We present the detailed solution method elsewhere (\cite{vdVen2003}),
and here we summarise the main results. 
In ongoing work we will apply our solutions, and will use them
together with the mean streaming motions 
(\cite{1994ApJ...425..458S}) 
to study the properties of the observed velocity and dispersion fields
of triaxial galaxies.


\section{The Jeans equations for separable models}
\label{sec:jeanseqnsseparablemodels}

We define confocal ellipsoidal coordinates ($\lambda,\mu,\nu$) as the
three roots for $\tau$ of 
\begin{equation}
  \label{eq:defellipsiodalcoord}
  \frac{x^2}{\tau+\alpha} + \frac{y^2}{\tau+\beta} +
  \frac{z^2}{\tau+\gamma} = 1 \; ,
\end{equation}
with ($x,y,z$) the usual Cartesian coordinates, and with constants
$\alpha,\beta$ and $\gamma$ such that $-\gamma
\leq \nu\leq -\beta \leq \mu \leq -\alpha \leq \lambda$.
Surfaces of constant $\lambda$ are ellipsoids, and surfaces of
constant $\mu$ and $\nu$ are hyperboloids of one and two sheets,
respectively. 
The confocal ellipsoidal coordinates are approximately Cartesian near
the origin and become conical at large radii, i.e., equivalent to
spherical coordinates. 

We consider models with a gravitational potential of St\"ackel form 
\begin{equation}
  \label{eq:formstackelpotential}
  V_S(\lambda,\mu,\nu) = 
  -\frac{F(\lambda)}{(\lambda\!-\!\mu)(\lambda\!-\!\nu)}
  -\frac{F(\mu)}{(\mu\!-\!\nu)(\mu\!-\!\lambda)}
  -\frac{F(\nu)}{(\nu\!-\!\lambda)(\nu\!-\!\mu)} \; , 
\end{equation}
where $F(\tau)$ is an arbitrary smooth function. 
This potential is the most general form for which the Hamilton--Jacobi
equation separates 
(\cite{1962MNRAS.124...95L}, 
 \cite{histStackel1890}) 
All orbits have three exact isolating integrals of motion, which are
quadratic in the velocities 
(e.g. \cite{1985MNRAS.216..273D}). 
There are no irregular orbits, so that Jeans' theorem is strictly
valid 
(\cite{1962MNRAS.124....1L}), 
and the distribution function $f$ is a function of the three integrals. 
Therefore, out of the six symmetric second-order velocity moments,
defined as  
\begin{equation}
  \label{eq:defvelmoments}
  \langle v_i v_j \rangle (\vec{x}) =
  \frac1\varrho \int\!\!\int\!\!\int 
  v_i v_j f(\vec{x},\vec{v}) \; \D ^3v \; , 
  \quad 
  (i,j=1,2,3),
\end{equation}
with density $\varrho$, the three mixed moments vanish, and we are
left with  $\langle v_\lambda^2 \rangle$, $\langle v_\mu^2 \rangle$
and $\langle v_\nu^2 \rangle$, related by three Jeans equations. 
These were first derived by Lynden--Bell 
(\cite{thesisLynden-Bell}), 
and can be written in the following form
(\cite{vdVen2003})
\begin{subeqnarray}
  \frac{\partial S_{\lambda\lambda}}{\partial \lambda} -
  \frac{S_{\mu\mu}}{2(\lambda\!-\!\mu)} -
  \frac{S_{\nu\nu}}{2(\lambda\!-\!\nu)} & = & g_1(\lambda,\mu,\nu) \; , 
  \label{eq:jeanstriaxial_lambdaS} \\
  \frac{\partial S_{\mu\mu}}{\partial \mu} -
  \frac{S_{\nu\nu}}{2(\mu\!-\!\nu)} -
  \frac{S_{\lambda\lambda}}{2(\mu\!-\!\lambda)} & = & g_2(\lambda,\mu,\nu) \; , 
  \label{eq:jeanstriaxial_muS} \\
  \frac{\partial S_{\nu\nu}}{\partial \nu} -
  \frac{S_{\lambda\lambda}}{2(\nu\!-\!\lambda)} -
  \frac{S_{\mu\mu}}{2(\nu\!-\!\mu)} & = & g_3(\lambda,\mu,\nu) \; , 
  \label{eq:jeanstriaxial_nuS}
\end{subeqnarray}
where we have defined the diagonal components of the stress tensor
\begin{equation}
  \label{eq:definitionStautau} 
  S_{\tau\tau}(\lambda,\mu,\nu) =
  \sqrt{ (\lambda\!-\!\mu) (\lambda\!-\!\nu) (\mu\!-\!\nu) } \, 
  \varrho\langle v_\tau^2 \rangle \; , 
  \qquad \tau=\lambda,\mu,\nu,
\end{equation}
and the functions $g_1$, $g_2$ and $g_3$ depend on the density
and potential (\ref{eq:formstackelpotential}) as
\begin{equation}
  \label{eq:deffunctiong_1}
  g_1(\lambda,\mu,\nu) = 
  - \sqrt{ (\lambda\!-\!\mu) (\lambda\!-\!\nu) (\mu\!-\!\nu) } \, 
  \varrho \, \frac{\partial V_S}{\partial \lambda} \; ,  
\end{equation}
where $g_2$ and $g_3$ follow from $g_1$ by cyclic permutation 
$\lambda\to\mu\to\nu\to\lambda$.
Similarly, the three Jeans equations follow from each other by
cyclic permutation. 
The stress components have to satisfy the following continuity
conditions 
\begin{equation}
  \label{eq:triaxialcontcond}
  S_{\lambda\lambda}(-\alpha,-\alpha,\nu) = 
  S_{\mu\mu}(-\alpha,-\alpha,\nu) \; ,
  \quad
  S_{\mu\mu}(\lambda,-\beta,-\beta) = 
  S_{\nu\nu}(\lambda,-\beta,-\beta) \; ,
\end{equation}
at the focal ellipse ($\lambda=\mu=-\alpha$) and focal hyperbola
($\mu=\nu=-\beta$), respectively. 

We prefer the form (\ref{eq:definitionStautau}) for the stresses
instead of the more common definition without the square 
root, since it results in more convenient and compact expressions.  
In self-consistent models, the density $\varrho$ equals $\varrho_S$,
with $\varrho_S$ related to $V_S$ by Poisson's equation.  
The Jeans equations, however, do not require self-consistency, so that
we make no assumptions on the form of $\varrho$ other than that it is
triaxial, i.e., a function of $(\lambda,\mu,\nu)$, and that it tends
to zero at infinity.  


\section{The two-dimensional case}
\label{sec:2Dcase}

When two or all three of the constants $\alpha$, $\beta$ or $\gamma$
in (\ref{eq:defellipsiodalcoord}) are equal, the triaxial St\"ackel
models reduce to limiting cases with more symmetry and thus with fewer
degrees of freedom.  
Solving the Jeans equations for oblate, prolate, elliptic disc and
scale-free models reduces to the same two-dimensional problem
(\cite{1989MNRAS.236..801E}, 
 \cite{2000MNRAS.318.1131E},  
 \cite{vdVen2003}), 
of which the simplest form is the pair of Jeans equations for
St\"ackel discs 
\begin{subeqnarray}
    \frac{\partial S_{\lambda\lambda}}{\partial \lambda} -
    \frac{S_{\mu\mu}}{2(\lambda\!-\!\mu)} & = & 
    g_1(\lambda,\mu) \; , 
    \label{eq:jeanselliptic_lambdaS} \\
    \frac{\partial S_{\mu\mu}}{\partial \mu} -
    \frac{S_{\lambda\lambda}}{2(\mu\!-\!\lambda)} & = & 
    g_2(\lambda,\mu) \; ,  
    \label{eq:jeanselliptic_muS}
\end{subeqnarray}
with at the foci ($\lambda=\mu=-\alpha$) the continuity condition
\begin{equation}
  \label{eq:disccontcond}
  S_{\lambda\lambda}(-\alpha,-\alpha) = S_{\mu\mu}(-\alpha,-\alpha) \; .
\end{equation}
In this case the stress components and the functions $g_1$ and $g_2$ are 
\begin{equation}
  \label{eq:defStt_disc}
  S_{\tau\tau}(\lambda,\mu) =
  \sqrt{ (\lambda\!-\!\mu) } \, 
  \varrho\langle v_\tau^2 \rangle 
  \quad (\tau=\lambda,\mu),
  \quad
  g_1(\lambda,\mu) = 
  - \sqrt{ (\lambda\!-\!\mu) } \, 
  \varrho \, \frac{\partial V_S}{\partial \lambda} \; ,  
\end{equation}
where $g_2$ follows from $g_1$ by interchanging
$\lambda\leftrightarrow\mu$, and $\varrho$ denotes a surface density. 

The two Jeans equations (8) 
can be recast into a single second-order partial differential equation
in either $S_{\lambda\lambda}$ or $S_{\mu\mu}$, which can be solved by
employing standard techniques  like Riemann's method 
(\cite{C75}, \cite{vdVen2003}).
However, these standard techniques do not carry over to the triaxial
case, and we therefore introduce an alternative method, based on the
superposition of singular solutions. 

\begin{figure}[t]
  \begin{center}
    \includegraphics[draft=false,scale=0.6,trim=2cm 4cm 0cm
    9cm]{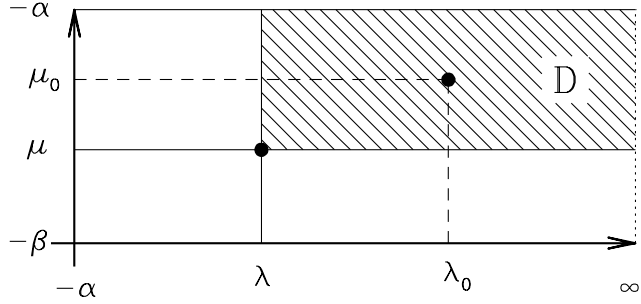} 
  \end{center}
  \caption[]{\slshape 
    The $(\lambda_0, \mu_0)$-plane. 
    The total stress at a field point $(\lambda,\mu)$ consists of the
    weighted contributions from source points at $(\lambda_0,\mu_0)$ 
    in the domain $D$. 
  }
  \label{fig:sourcepointcontribution}
\end{figure}

We consider a simpler form of (8) 
by substituting for $g_1$ and $g_2$, respectively $\tilde{g}_1=0$ and
$\tilde{g}_2 = \delta(\lambda_0\!-\!\lambda) \delta(\mu_0\!-\!\mu)$.    
We refer to solutions of these simplified Jeans equations as
\textit{singular solutions}.
Singular solutions can be interpreted as contributions to the stresses 
at a fixed field point $(\lambda,\mu)$ due to a source point in 
$(\lambda_0,\mu_0)$ (Fig.~\ref{fig:sourcepointcontribution}).   
The full stress at the field point can be obtained by adding all
source point contributions, each with a weight that depends on the
local density and potential.
Once we know the singular solutions, we can use the superposition
principle to construct the the solution of the full Jeans equations
(8).

Since the derivative of a step-function $\mathcal{H}$ is equal to a
delta-function, it follows that the singular solutions must have the
form  
\begin{eqnarray} 
  \label{eq:functionalformsSllandSmm}
  S_{\lambda\lambda} 
  & = & 
  A(\lambda,\mu) \mathcal{H}(\lambda_0\!-\!\lambda)
  \mathcal{H}(\mu_0\!-\!\mu) \; , 
  \nonumber \\*[-5pt] \\*[-5pt] 
  S_{\mu\mu}
  & = & 
  B(\lambda,\mu) \mathcal{H}(\lambda_0\!-\!\lambda)
  \mathcal{H}(\mu_0\!-\!\mu) \!-\! \delta(\lambda_0\!-\!\lambda)
  \mathcal{H}(\mu_0\!-\!\mu) \; .  
  \nonumber  
\end{eqnarray}
where the functions $A$ and $B$ must solve the homogeneous Jeans
equations, i.e., (8) 
with zero right-hand side, and satisfy the following boundary
conditions   
\begin{equation}
  \label{eq:dischombc1and2}
  A(\lambda_0,\mu) = \frac{1}{2(\lambda_0\!-\!\mu)} \; , 
  \quad 
  B(\lambda,\mu_0) = 0 \; .
\end{equation}
We solve this two-dimensional homogeneous boundary problem by
superposition of particular solutions.  We first derive a particular
solution of the homogeneous Jeans equations with a free parameter $z$,
which we assume to be complex.  We then construct a linear combination
of these particular solutions by integrating over $z$.  We choose the
integration contours in the complex $z$-plane, such that the boundary
conditions (\ref{eq:dischombc1and2}) are satisfied simultaneously.
The resulting homogeneous solutions are complex contour integrals,
which can be evaluated in terms of the complete elliptic integral of
the second kind, $E(w) \equiv \int_0^\frac{\pi}{2} \D \theta \,
\sqrt{1-w\sin^2\theta}$, and its derivative $E'(w)$, as
\begin{equation}
  \label{eq:homogeneoussolutions}
  A = \frac{ E(w) }{ \pi(\lambda_0\!-\!\mu)} \; , 
  \quad
  B = -\frac{ 2w E'(w) }{ \pi (\lambda_0\!-\!\lambda) } \; ,
  \quad 
  \mathrm{with}
  \quad
   w = \frac{(\lambda_0\!-\!\lambda) (\mu_0\!-\!\mu) }{
  (\lambda_0\!-\!\mu_0) (\lambda\!-\!\mu) } \; .
\end{equation}
We obtain a similar system of simplified Jeans equations by
interchanging the expressions for $\tilde{g}_1$ and $\tilde{g}_2$. 
The singular solutions of this simplified system follow from
(\ref{eq:functionalformsSllandSmm}) by interchanging
$\lambda\leftrightarrow\mu$ and $\lambda_0\leftrightarrow\mu_0$ at the
same time. 

To find the solution to the full Jeans equations
(8) 
at $(\lambda,\mu)$, we multiply the latter singular solutions and
(\ref{eq:functionalformsSllandSmm}) by  $g_1(\lambda_0,\mu_0)$ and
$g_2(\lambda_0,\mu_0)$ respectively, and integrate over 
$D=\{(\lambda_0,\mu_0)$: $\lambda\le\lambda_0\le\infty,
\mu\le\mu_0\le-\alpha \}$ (Fig. \ref{fig:sourcepointcontribution}). 
This gives the first two integrals of the two equations
(\ref{eq:discsolSll}) and (\ref{eq:gensolSmm}) below. 
The remaining terms are due to the non-vanishing stress at the
boundary $\mu=-\alpha$, and are found by multiplying the singular
solutions (\ref{eq:functionalformsSllandSmm}), evaluated at
$\mu_0=-\alpha$, by $-S_{\mu\mu}(\lambda_0,-\alpha)$ and integrating
over $\lambda_0$ in $D$.    
The final result for the solution of the Jeans equations
(8) 
for St\"ackel discs, after using
the evaluations (\ref{eq:homogeneoussolutions}), is 
\begin{subeqnarray}
  S_{\lambda\lambda}(\lambda,\mu) = 
  \int\limits_\lambda^\infty \hspace{-4pt} \D \lambda_0 
  \hspace{-4pt} \int\limits_\mu^{-\alpha} \hspace{-4pt} \D \mu_0
  \biggl[
  -g_1(\lambda_0,\mu_0)\frac{2wE'(w)}{\pi(\mu_0\!-\!\mu)} 
  \!+\!g_2(\lambda_0,\mu_0)\frac{E(w)}{\pi(\lambda_0\!-\!\mu)}
  \biggr] 
  \nonumber \\ 
  -\int\limits_\lambda^\infty \hspace{-4pt} \D \lambda_0 
  \; g_1(\lambda_0,\mu) 
  - \int\limits_\lambda^\infty \hspace{-4pt} \D \lambda_0 
  \, S_{\mu\mu}(\lambda_0,-\alpha) \, 
  \biggl[ \frac{E(w)}{\pi(\lambda_0\!-\!\mu)} 
  \biggr]_{\mu_0=-\alpha} \hspace{-20pt}, 
  \label{eq:discsolSll} \\
  \vspace{-10pt}
  S_{\mu\mu}(\lambda,\mu) = 
  \int\limits_\lambda^\infty \hspace{-4pt} \D \lambda_0 
  \hspace{-4pt} \int\limits_\mu^{-\alpha} \hspace{-4pt} \D \mu_0
  \biggl[
  -g_1(\lambda_0,\mu_0)\frac{E(w)}{\pi(\lambda\!-\!\mu_0)} 
  \!-\!g_2(\lambda_0,\mu_0)\frac{2wE'(w)}{\pi(\lambda_0\!-\!\lambda)}
  \biggr] \nonumber \\ 
  -\int\limits_\mu^{-\alpha} \hspace{-4pt} \D \mu_0
  \; g_2(\lambda,\mu_0) 
  + S_{\mu\mu}(\lambda,-\alpha)
  - \int\limits_\lambda^\infty \hspace{-4pt} \D \lambda_0
  \, S_{\mu\mu}(\lambda_0,-\alpha) 
  \biggl[-\frac{2wE'(w)}{\pi(\lambda_0\!-\!\lambda)} 
  \biggr]_{\mu_0=-\alpha} \hspace{-20pt}.
  \label{eq:gensolSmm}
\end{subeqnarray}
The solution depends on $\varrho$ and $V_S$ through $g_1$ and $g_2$.
This means that, for given $\varrho$ and $V_S$, the solution is uniquely
determined once we have prescribed $S_{\mu\mu}$ at the boundary
$\mu=-\alpha$.  
At this boundary, $S_{\lambda\lambda}$ is related to $S_{\mu\mu}$ by
the first Jeans equation (\ref{eq:jeanselliptic_lambdaS}), evaluated
at $\mu=-\alpha$, up to an integration constant, which is fixed by the
continuity condition (\ref{eq:disccontcond}). 
We are thus free to specify either of the two stress components at
$\mu=-\alpha$. 


\section{The general case}
\label{sec:generalcase}

The singular solution method introduced in the previous section can be
extended to three dimensions to solve the Jeans equations
(4) 
for triaxial St\"ackel models. 
Although the calculations are more complex for a triaxial model, the
stepwise solution method is similar to that in two dimensions. 

We simplify the Jeans equations (4) 
by setting two of the three functions $g_1$, $g_2$ and $g_3$ to
zero and the remaining equal to $\delta(\lambda_0\!-\!\lambda)
\delta(\mu_0\!-\!\mu) \delta(\nu_0\!-\!\nu)$. 
In this way, we obtain three similar simplified systems ($i=1,2,3$),
each with three singular solutions 
$S_i^{\tau\tau}(\lambda,\mu,\nu;\lambda_0,\mu_0,\nu_0)$ 
($\tau=\lambda,\mu,\nu$), that describe the stress components at a
fixed field point $(\lambda,\mu,\nu)$ due to a source point in
$(\lambda_0,\mu_0,\nu_0)$.   

The singular solutions have a form that is similar to that in the
two-dimensio\-nal case (\ref{eq:functionalformsSllandSmm}). 
They consist of combinations of step-functions and delta-functions
multiplied by functions that are the solutions of homogeneous boundary
problems.  
The functions that must solve a two-dimensional homogeneous boundary
problem can be found as in \S \ref{sec:2Dcase}, and can be expressed
in terms of complete elliptic integrals, 
cf. (\ref{eq:homogeneoussolutions}).
The singular solutions in the general case also contain three
functions $A$, $B$ and $C$ that must solve the triaxial homogeneous
Jeans equations, i.e., (4) 
with zero right-hand side, and satisfy three boundary conditions. 
This three-dimensional homogeneous boundary problem can be solved by
integra\-ting a \textit{two}-parameter particular solution over both
its complex parameters, and choosing the combination of contours such
that the three boundary conditions are satisfied simultaneously.  
The resulting homogeneous solutions $A$, $B$ and $C$ are products
of complex contour integrals, and can be evaluated as sums of products
of complete \textit{hyper}elliptic integrals.  

To find the solution of the full Jeans equations
(4) 
, we multiply each singular solution
$S_i^{\tau\tau}$ by $g_i(\lambda_0,\mu_0,\nu_0)$, so that the
contribution from the source point naturally depends on the local
density and potential. 
Then, for each coordinate $\tau=\lambda,\mu,\nu$, we add the three
weighted singular solutions, and integrate over a finite volume
within the physical region 
$-\gamma\le\nu\le-\beta\le\mu\le-\alpha\le\lambda$.
This results in the following general solution of the Jeans equations  
(4) 
for triaxial St\"ackel models
\begin{eqnarray}
  \label{eq:generaltriaxsolStt}
  S_{\tau\tau} 
  (\lambda,\mu,\nu) & = &
  \int\limits_\lambda^{\lambda_e} \hspace{-4pt} \D \lambda_0 
  \int\limits_\mu^{\mu_e} \hspace{-4pt} \D \mu_0  
  \int\limits_\nu^{\nu_e} \hspace{-4pt} \D \nu_0
  \sum_{i=1}^3 g_i(\lambda_0,\mu_0,\nu_0)
  \, S_i^{\tau\tau} (\lambda,\mu,\nu;\lambda_0,\mu_0,\nu_0) 
  \nonumber \\*[-3pt]
  & & -
  \int\limits_\mu^{\mu_e} \hspace{-4pt} \D \mu_0 
  \int\limits_\nu^{\nu_e} \hspace{-4pt} \D \nu_0  
  \, S_{\lambda\lambda}(\lambda_e,\mu_0,\nu_0) 
  \, S_1^{\tau\tau} (\lambda,\mu,\nu;\lambda_e,\mu_0,\nu_0)  
  \nonumber \\*[-3pt]
  & & -
  \int\limits_\nu^{\nu_e} \hspace{-4pt} \D\nu_0    
  \int\limits_\lambda^{\lambda_e} \hspace{-4pt} \D \lambda_0 
  \, S_{\mu\mu}(\lambda_0,\mu_e,\nu_0) 
  \, S_2^{\tau\tau}
  (\lambda,\mu,\nu;\lambda_0,\mu_e,\nu_0)  
  \nonumber \\*[-3pt]
  & & -
  \int\limits_\lambda^{\lambda_e} \hspace{-4pt} \D \lambda_0 
  \int\limits_\mu^{\mu_e} \hspace{-4pt} \D \mu_0    
  \, S_{\nu\nu}(\lambda_0,\mu_0,\nu_e)
  \, S_3^{\tau\tau} (\lambda,\mu,\nu;\lambda_0,\mu_0,\nu_e) \; ,  
\end{eqnarray}
with $\tau=\lambda, \mu, \nu$.
Whereas the integration limits $\lambda$, $\mu$ and $\nu$ are fixed
due to the position of the field point, the limits $\lambda_e$,
$\mu_e$ and $\nu_e$ are not, and may be any value in the
corresponding physical ranges, i.e., $\lambda_e\in[-\alpha,\infty]$,
$\mu_e\in[-\beta,-\alpha]$ and $\nu_e\in[-\gamma,-\beta]$, but
$\lambda_e \ne -\alpha$.  
The latter choice would lead to solutions which generally have the
incorrect radial fall-off, and hence are non-physical.  
If we choose $\lambda_e=\infty$, there is no contribution from the
second line in (\ref{eq:generaltriaxsolStt}) due to vanishing stress
at large distance.
If we furthermore take $\mu_e=-\alpha$ and $\nu_e=-\beta$, the
integration volume becomes the three-dimensional extension of $D$
(Fig. \ref{fig:sourcepointcontribution}). 

Whereas the volume integral in (\ref{eq:generaltriaxsolStt}) already
solves the inhomogeneous Jeans equations (4) 
, the three area integrals are needed to obtain the correct values at
the boundary surfaces $\lambda=\lambda_e$, $\mu=\mu_e$ and
$\nu=\nu_e$.  On each of these surfaces the three stress components
are related by two of the three Jeans equations
(4) 
and the continuity conditions
(\ref{eq:triaxialcontcond}).  Since the (weight) functions $g_i$ are
known for given $\varrho$ and $V_S$, this means that the solution
(\ref{eq:generaltriaxsolStt}) yields all three stresses everywhere in
the triaxial model, once one of the stress components is prescribed on
the three boundary surfaces.  If we take $\lambda_e=\infty$ and
$\mu_e=\nu_e=-\beta$, the contributing boundary surfaces reduce to the
single $(x,z)$-plane, containing the long and the short axis of the
galaxy.  This compares well with Schwarzschild
(\cite{1993ApJ...409..563S}), who used the same plane to start his
numerically calculated orbits from.


\section{Discussion and conclusions}
\label{sec:discconc}

Eddington (\cite{1915MNRAS..76...37E}) 
showed that the velocity ellipsoid in a triaxial galaxy with a
separable potential of St\"ackel form is everywhere aligned with the
confocal ellipsoidal coordinate system in which the equations of
motion separate. 
Lynden--Bell (\cite{thesisLynden-Bell}) 
derived the three Jeans equations which relate the three principal
stresses to the potential and the density. 
Solutions were found for the various two-dimensional limiting cases,
but with methods that do not carry over to the general case, which
remained unsolved. 
We have presented an alternative solution method, based on
the superposition of singular solutions (see \cite{vdVen2003} for
details). 
This approach, unlike the standard techniques, can be generalised to
solve the three-dimensional system.     
The resulting solutions contain complete (hyper)elliptic integrals,
which can be evaluated in a straightforward way. 

The general Jeans solution is not unique, but requires specification
of principal stresses at certain boundary surfaces, given a separable
triaxial potential and a triaxial density distribution (not
necessarily the one that generates the potential). 
These boundary surfaces can be taken to be the plane containing the
long and the short axis of the galaxy, and, more specifically, the
part that is crossed by all three families of tube orbits and the box
orbits.  

The set of all Jeans solutions (\ref{eq:generaltriaxsolStt}) contains
all the stresses that are associated with the physical distribution 
functions $f \geq 0$, but, as in the case of spherical and 
axisymmetric models, also contains solutions which are
unphysical, e.g., those associated with distribution functions that
are negative in some parts of phase space. 
The many examples of the use of spherical and axisymmetric Jeans
models in the literature suggest nevertheless that the Jeans solutions
can be of significant use.

While triaxial models with a separable potential do not provide an
adequate description of the nuclei of galaxies with cusped luminosity
profiles and a massive central black hole
(\cite{1986MNRAS.221.1001D}), 
they do catch much of the orbital structure at larger radii, and in
some cases even provide a good approximation of the galaxy potential.
The solutions for the mean streaming motions, i.e., the first velocity
moments of the distribution function, are helpful in
understanding the variety of observed velocity fields in giant
elliptical galaxies and constraining their intrinsic shapes 
(e.g. 
\cite{1994MNRAS.271..924A},  
\cite{2001AJ....121..244S}, 
\cite{1999AJ....117..126S}).  
We expect that the projected velocity dispersion fields that can be
derived from our Jeans solutions will be similarly useful, and, in
particular, that they can be used to establish which combinations of
viewing directions and intrinsic axis ratios are firmly ruled out by
the observations. 

It is remarkable that the entire Jeans solution can be written down by
means of classical methods. 
This suggests that similar solutions can be found for the higher
dimensional analogues of (4) 
, most likely involving hyperelliptic integrals of higher order. 
It is also likely that the higher-order velocity moments for the
separable triaxial models can be found by similar analytic means, but
the effort required may become prohibitive.

%

\end{document}